\documentclass[twoside]{dis08}
\usepackage[latin1]{inputenc}
\usepackage[dvips]{graphicx,epsfig,color}
\usepackage{wrapfig,rotating}
\usepackage{amssymb,amsmath,array}

\begin{document}
\title{Recent PHENIX Results on $\pi^\pm$ and $\eta$ Production in polarized pp Collisions at RHIC at $\sqrt{s}=200$~GeV}
%***********************************************************************

% AUTHORS INFORMATION AREA

%***********************************************************************

\author{Frank Ellinghaus$^1$ \thanks{This work is supported in part by the US Department of Energy.}
\vspace{.3cm}\\
University of Colorado, Boulder, Colorado 80309, USA \\
E-mail: Frank.Ellinghaus@desy.de
\vspace{.1cm}\\
(for the PHENIX Collaboration)}
%***********************************************************************

% END OF AUTHORS INFORMATION AREA

%***********************************************************************

\maketitle

\begin{abstract}
Measurements of double helicity asymmetries for inclusive
hadron production in polarized proton-proton collisions are
sensitive to spin--dependent parton distribution functions, in particular
to the gluon distribution, $\Delta g$.
This study focuses on double helicity asymmetries in $\eta$ ($\vec{p}+\vec{p} \rightarrow
\eta+X$) and charged pion production
($\vec{p}+\vec{p} \rightarrow \pi^\pm+X$) at mid-rapidity, as well as on the $\eta$ cross section
and fragmentation functions.

\end{abstract}

\section{Introduction}
Present knowledge about spin--dependent parton distribution functions (PDFs)
in the nucleon mainly comes from next-to-leading order (NLO) QCD fits
(see, e.g., \cite{bb,deflorian}) to the spin--dependent structure function $g_1$
as measured in polarized inclusive deep--inelastic scattering (DIS) experiments (see, e.g., \cite{hermes_g1,compass_g1_d}).
The resulting spin--dependent PDFs for the gluon have rather large uncertainties due to
the fact that the exchanged virtual photon does not couple directly, i.e., at
leading order, to the gluon. Thus additional data from polarized pp
scattering, in which longitudinally polarized gluons are directly probed via scattering off
longitudinally polarized gluons or quarks, should greatly reduce the
uncertainties in the gluon distribution. 
%in the NLO fits. 
This has recently been demonstrated in a global NLO fit \cite{DSSV} using, for the first
time, the available inclusive and semi-inclusive DIS data together with 
first results from polarized pp scattering 
%from the PHENIX and STAR collaborations 
at the Relativistic Heavy Ion Collider (RHIC).
The results used were double helicity asymmetries in inclusive $\pi^0$ \cite {allpi0} 
and jet \cite{star_jets} production from the PHENIX and STAR experiments, respectively.
The double helicity asymmetry is defined as
\begin{equation} %\nonumber
A_{LL} = \frac{\sigma^{++}-\sigma^{+-}}{\sigma^{++}+\sigma^{+-}}
= \frac{\Delta \sigma}{\sigma}, \quad \text{with} \quad
\Delta \sigma
%\propto \sum_{abc}\Delta f_a(x_1)\otimes \Delta f_b(x_2)\otimes\Delta\hat{\sigma}^{ab\rightarrow cX'}(\hat{s})\otimes D^{h}_c
\propto \sum_{abc}\Delta f_a\otimes \Delta f_b\otimes\Delta\hat{\sigma}^{ab\rightarrow cX'}\otimes D^{h}_c,
\label{cross_sec_asy}
\end{equation}
where the cross section $\sigma^{++}$ ($\sigma^{+-}$)
describes the reaction where both protons have the same (opposite) helicity.
The spin--dependent term is given on the rhs of Eqn. \ref{cross_sec_asy},
where $\Delta f_a$, $\Delta f_b$ represent the spin--dependent PDFs for quarks (u,d,s) and
gluons, and $\Delta \hat{\sigma}$ are the spin--dependent hard scattering cross
sections calculable in perturbative QCD.
The fragmentation functions (FFs)
$D^h_c$ represent the probability for a certain parton $c$ to fragment into a certain
hadron $h$, and thus they are not needed in the case of jet production.

This study focuses on $A_{LL}$ in $\eta$ ($\vec{p}+\vec{p} \rightarrow
\eta+X$) and charged pion production
($\vec{p}+\vec{p} \rightarrow \pi^\pm+X$), as well as on the $\eta$ cross section
and fragmentation functions.

\section{Eta cross section and fragmentation functions}
%For the measurement of the double helicity asymmetry,
The $\eta$ meson is
reconstructed via its main decay channel $\eta \rightarrow  \gamma \gamma$ with a branching
ratio of about 40\%. The three-body decay $\eta \rightarrow \pi^+ \pi^- \pi^0$
not only has a smaller branching ratio of about 23\%, but in addition also has
a smaller acceptance in the PHENIX spectrometer \cite{phenix} and therefore has not been
considered.
%The cross section data for the $\eta$ presented here were taken at the PHENIX \cite{phenix} experiment 
%in 2005. 
The primary detector used in this analysis is the electromagnetic calorimeter,
%which is
located at a radial distance of about 5~m from the beam pipe.
It covers the pseudo--rapidity range $|\eta| < 0.35$ and has an azimuthal
acceptance of $\Delta \phi = \pi$.

\begin{figure}
\begin{center}
\includegraphics[width=0.62\textwidth]{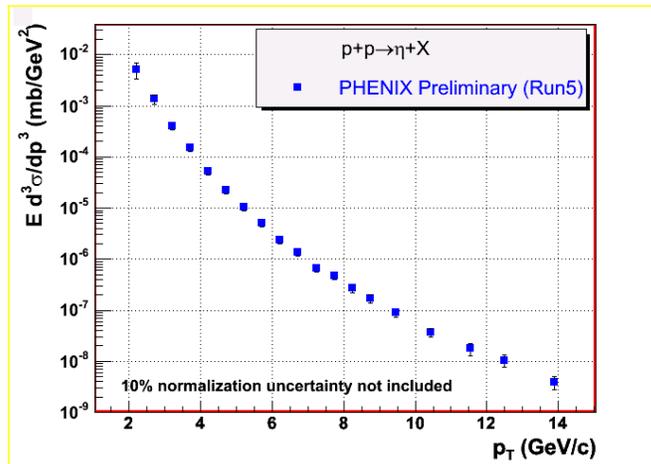}
\end{center}
\caption{Cross section for mid-rapidity inclusive $\eta$ production at $\sqrt{s}=200$~GeV as a function of $p_T$.}
\label{eta_cs}
\end{figure}

\begin{figure}
\includegraphics[width=0.495\columnwidth]{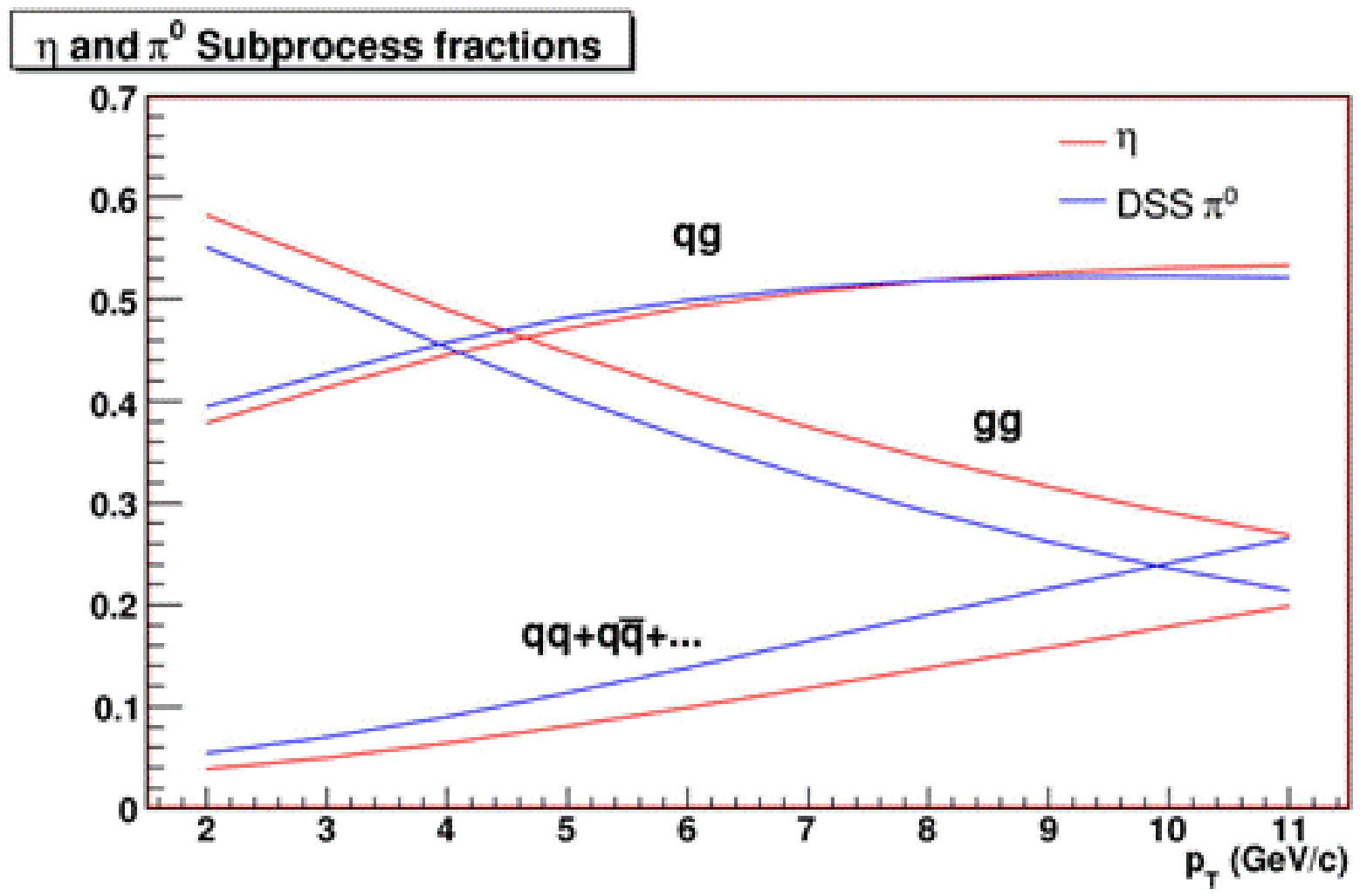}
\includegraphics[width=0.495\columnwidth]{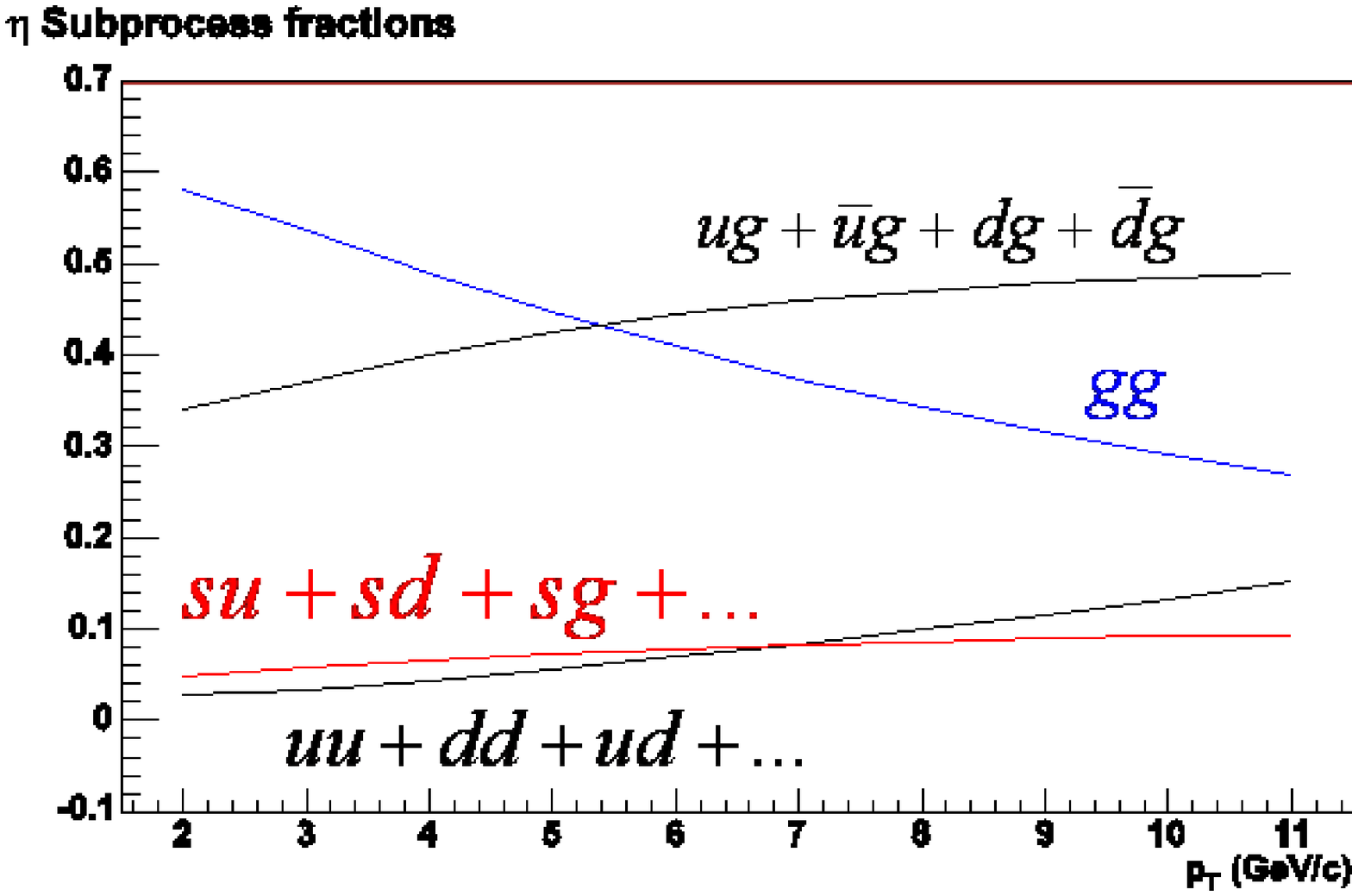}
\caption{Left panel: Fractional contribution of gluon--gluon ($gg$), quark--gluon($qg$), and
  quark--quark ($qq$) scattering to $\eta$ (preliminary) and $\pi^0$ \cite{DSS}
  production as a function of $p_T$.
Right panel: The fractional contributions to $\eta$ production (preliminary) are
  shown for scattering involving (red line) or not
  involving (black lines and blue line) strange quarks.}
\label{subprocess_frac}
\end{figure}

The preliminary $\eta$ cross section as a function of
$p_T$ from data taken in 2005 is shown in Fig.~\ref{eta_cs}.
It is consistent with an earlier PHENIX cross section measurement
\cite{eta_long} covering a smaller range in $p_T$. % from 2.75 to 11 GeV. 
Using the code of Ref.~\cite{DSS} this result from $pp$ scattering together with
various $\eta$ cross section 
measurements from $e^+e^-$ scattering have been used to extract preliminary fragmentation functions for
$\eta$ production. The resulting contributions of the various scattering subprocesses
gluon--gluon ($gg$), quark--gluon($qg$), and quark--quark ($qq$) to the $\eta$ 
production as a function of $p_T$ are shown in Fig. \ref{subprocess_frac}. 
They are compared to the fractional contribution of the subprocesses to the
$\pi^0$ production 
%based on Ref. 
\cite{DSS}. The $\eta$ production has a larger contribution from $gg$ scattering and
thus enhanced sensitivity to the gluon PDF when compared to the $\pi^0$. This is
expected due to the additional strange quark contribution in the wave function of
$\eta$ mesons which is absent for the $\pi^0$. 

The $s$-quark contribution in the $\eta$ wave function
also opens up the possibility of studying the polarized strange
quark PDF ($\Delta s$). Special interest arises from the fact that $\Delta s$
has been assumed to be
negative based on inclusive DIS data, however, semi-inclusive DIS
measurements \cite{hermes_semi, hermes_deltas} find $\Delta s$ to be compatible with zero or slightly
positive. The $\eta$ $A_{LL}$ measurement is in principle sensitive to $\Delta s$ as
shown in Fig. \ref{subprocess_frac}, where the contribution of subprocesses involving $s$ quarks rises
to 10\% at $p_T = 10$~GeV. However, 
%apart from the fact that this is a preliminary result it is also afflicted with 
potentially large uncertainties are possible due to the absence
of semi-inclusive data on $\eta$ production in this preliminary FF determination.

\section{Double Helicity Asymmetry for $\pi^\pm$ and $\eta$}
Experimentally, the double helicity
asymmetry (Eqn.~\ref{cross_sec_asy}) translates into
\begin{equation} %\nonumber
A_{LL} = \frac{1}{|P_B||P_Y|}\frac{N_{++}-RN_{+-}}{N_{++}+RN_{+-}}, \quad
\text{with} \quad 
R\equiv\frac{L_{++}}{L_{+-}},
\end{equation}
where $N_{++}$ ($N_{+-}$) 
%and $L_{++}$ ($L_{+-}$) 
is the experimental yield for the case where the beams
have the same (opposite) helicity.
The relative luminosity $R$ is measured by a coincident signal in two beam--beam counters, which 
have a full azimuthal coverage at a distance of about $\pm 1.4$~m from the target.
The achieved uncertainty on $R$ is on the order of $10^{-4}$.
The polarizations of the two colliding beams at RHIC are denoted by $P_B$ and $P_Y$. 
%They are measured by a proton-Carbon and by a hydrogen gas jet 
The degree of polarization is determined from the combined information of a
$\vec {p}C$ polarimeter~\cite{pC}, using an unpolarized ultra--thin
carbon ribbon target, 
and from $\vec {p}\vec{p}$ scattering, using a polarized atomic
hydrogen gas-jet target~\cite{jet}. 
The average polarization value for the data from 2005 (2006) is
$49$\% (57\%). 
%with an uncertainty of $20$\% (8.3\%) per beam, leading to a 40\% (16.6\%) scale uncertainty
%in the preliminary $A_{LL}$ result.

\begin{figure}
\includegraphics[width=0.495\columnwidth, height=5.8cm]{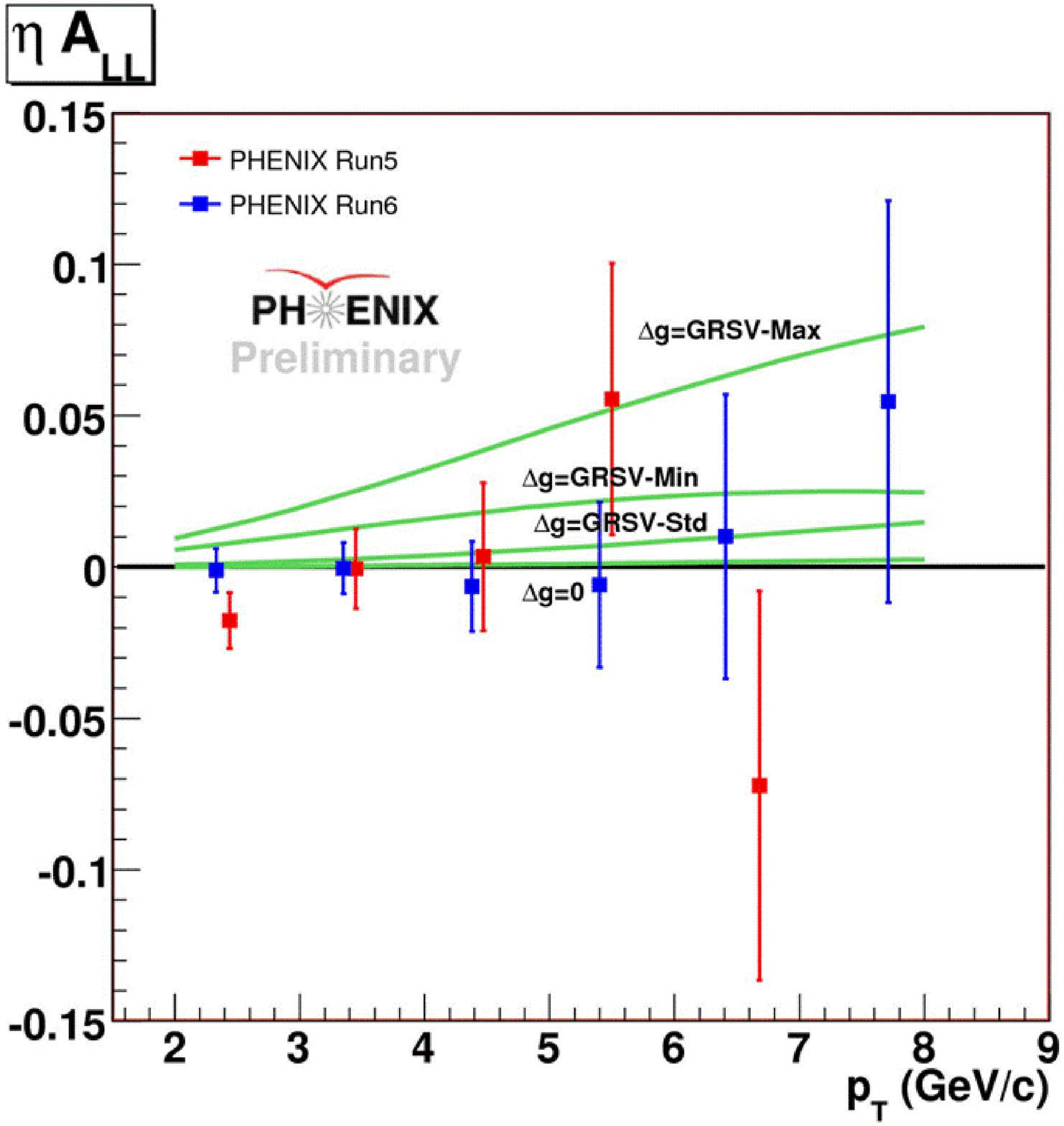}
\includegraphics[width=0.495\columnwidth]{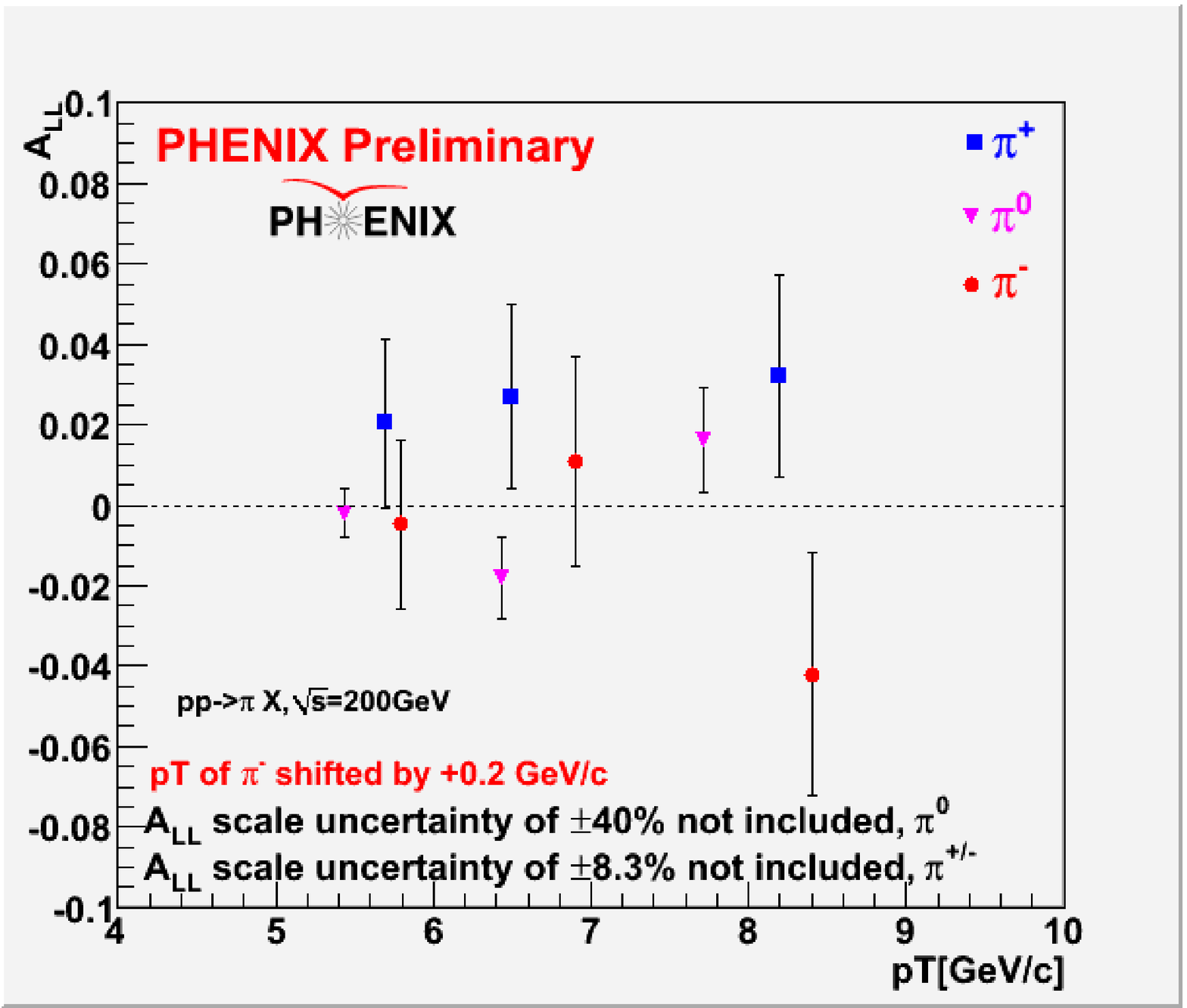}
\caption{Double helicity asymmetry for mid-rapidity inclusive $\eta$ (left
  panel) production from the 2005 \cite{elli_spin06} and 2006 data and pion (right
  panel) production from the 2006 data at $\sqrt{s}=200$~GeV as a
  function of $p_T$. The results for the $\eta$ asymmetry are compared to NLO pQCD
  calculations \cite{marco,GRSV}. See text for details.}
\label{asys}
\end{figure}

The double helicity asymmetry for $\eta$ production as
a function of $p_T$ from the 2005 \cite{elli_spin06} and 2006 data is shown in the left panel of Fig.~\ref{asys}.
It is consistent with zero over the measured range.
The results are compared to NLO pQCD calculations \cite{marco}, using
the above mentioned preliminary FFs and the GRSV set of polarized PDFs \cite{GRSV}.
The standard GRSV set ($\Delta g =$ GRSV-Std) has been modified by assuming the polarized gluon PDF
to be zero ($\Delta g = 0$), equal ($\Delta g =$ GRSV-Max), or opposite
($\Delta g =$ GRSV-Min) to the unpolarized gluon PDF at the input scale. 
It is apparent that the maximum and minimum scenarios are ruled out by this
data alone, and that future data can further constrain the polarized
gluon PDF.

The double helicity asymmetry for charged pion production as
a function of $p_T$ is shown in the right panel of Fig.~\ref{asys}.
Similar to the case of $\eta$ production the data has been compared to
NLO pQCD calculations \cite{url}. So far only the maximum gluon scenario can be
excluded, however, the advantage of the charged pion data is in its increased
sensitivity to the sign of $\Delta G$. 
In fact, the sign can be studied in a model independent way.
Due to the fact that $qg$ scattering dominates in the measured $p_T$ region and
that the polarized $u$ ($d$) quark PDFs are well known from DIS, a positive or
vanishing gluon polarization would lead to $A_{LL}^{\pi^+} > A_{LL}^{\pi^0}  >
A_{LL}^{\pi^-}$ due to
the increasing (decreasing) contribution of negatively (positively) polarized
$d$ ($u$) quarks. As can be seen in the right panel of Fig. \ref{asys}, in
which the PHENIX $A_{LL}$ result for $\pi^0$ production in the $p_T$ region of
interest is shown as well,
additional data will be needed to provide the desired constraint
on the sign of the polarized gluon PDF.

\begin{footnotesize}

\end{footnotesize}

% ****************************************************************************

% END OF BIBLIOGRAPHY AREA

% ****************************************************************************

\end{document}